%% file: 00-Main.tex
\documentclass[conference]{IEEEtran}
\IEEEoverridecommandlockouts
\usepackage{cite}
\usepackage{amsmath,amssymb,amsfonts}
\usepackage{algorithmic}
\usepackage{graphicx}
\usepackage{textcomp}
\usepackage{xcolor}
\usepackage{subcaption}
\usepackage{url}
\usepackage{booktabs}
\usepackage[normalem]{ulem}  

\usepackage[letterspace=-18,tracking=true,final]{microtype}

\def\BibTeX{{\rm B\kern-.05em{\sc i\kern-.025em b}\kern-.08em
		T\kern-.1667em\lower.7ex\hbox{E}\kern-.125emX}}
	
\def\fixme#1{\bgroup \color{red}{[{#1}]}\egroup}

\newcommand{\ib}[1]{\textcolor{blue}{[IB: \emph{#1}]}}

\begin{document}
	
	\title{DRAM Errors and Cosmic Rays:\\ Space Invaders or Science Fiction?}

	\makeatletter
	\newcommand{\linebreakand}{%
		\end{@IEEEauthorhalign}
		\hfill\mbox{}\par
		\mbox{}\hfill\begin{@IEEEauthorhalign}
	}
	\makeatother
	
	\author{		
		\IEEEauthorblockN{Isaac Boixaderas}
		\IEEEauthorblockA{
			\textit{Barcelona Supercomputing Center}\\
			isaac.boixaderas@bsc.es}
	
		\and
		
		\IEEEauthorblockN{Jorge Amaya}
		\IEEEauthorblockA{
			\textit{\ \ \ \ \ European Space Agency\ \ \ \ \ \ \ }\\
			jorge.amaya@esa.int}
		
		\and
		
		\IEEEauthorblockN{Sergi Mor\'e}
		\IEEEauthorblockA{
			\textit{Barcelona Supercomputing Center}\\
			sergi.more@bsc.es}
						
		\linebreakand
		
		\IEEEauthorblockN{Javier Bartolome}
		\IEEEauthorblockA{
			\textit{Barcelona Supercomputing Center}\\
			javier.bartolome@bsc.es}
		
		\and
		
		\IEEEauthorblockN{David Vicente}
		\IEEEauthorblockA{
			\textit{Barcelona Supercomputing Center}\\
			david.vicente@bsc.es}
		
		\and
		
		\IEEEauthorblockN{Osman Unsal}
		\IEEEauthorblockA{
			\textit{Barcelona Supercomputing Center}\\
			osman.unsal@bsc.es}
		
		\linebreakand
		
		\IEEEauthorblockN{Dimitris Gizopoulos}
		\IEEEauthorblockA{
			\textit{\ \ \ \ \ \ \ University of Athens\ \ \ \ \ \ \ \ \ }\\
			dgizop@di.uoa.gr}
	
		\and
		
		\IEEEauthorblockN{Paul M. Carpenter}
		\IEEEauthorblockA{
			\textit{Barcelona Supercomputing Center}\\
			paul.carpenter@bsc.es}
	
		\and
		
		\IEEEauthorblockN{Petar Radojkovi\'c}
		\IEEEauthorblockA{
			\textit{Barcelona Supercomputing Center}\\
			petar.radojkovic@bsc.es}
	
		\linebreakand
		
		\IEEEauthorblockN{Eduard Ayguad\'e}
		\IEEEauthorblockA{\textit{Barcelona Supercomputing Center} \\
			\textit{Universitat Politècnica de Catalunya}\\
			eduard.ayguade@bsc.es}
	
	\vspace{-0.3em}
	}
	
	\maketitle
	
	\thispagestyle{plain}
	\pagestyle{plain}

	\begin{abstract}
		It is widely accepted that cosmic rays are a plausible cause of DRAM errors in high-performance computing~(HPC) systems, 
		and various studies suggest that they could explain some aspects of the observed DRAM error behavior. 
		However, this phenomenon is insufficiently studied in production environments.
		
		We analyze the correlations between cosmic rays and DRAM errors on two HPC clusters: a production supercomputer with server-class DDR3-1600 and a prototype with LPDDR3-1600 and no hardware error correction. Our error logs cover 2000\,billion\,MB-hours for the MareNostrum~3 supercomputer and 135\,million\,MB-hours for the Mont-Blanc prototype. Our analysis combines quantitative analysis, formal statistical methods and machine learning.
		
		We detect no indications that cosmic rays have any influence on the DRAM errors.  
		To understand whether the findings are specific to systems under study, located at 100\,meters above the sea level, 
		the analysis should be repeated on other HPC clusters, especially the ones located on higher altitudes.  
		Also, analysis can (and should) be applied to revisit and extend numerous previous studies which use
		cosmic rays as a hypothetical explanation for some aspects of
		the observed DRAM error behaviors.  

	\end{abstract}
	
	\begin{IEEEkeywords}
		Memory system, Reliability, Cosmic rays
	\end{IEEEkeywords}
	
	\section{Introduction}
	\label{sec:Introduction}

\input{10-Introduction.tex}

	\section{Cosmic rays}
	\label{sec:CosmicRays}

\input{15-Background.tex}

	\section{MareNostrum~3 DRAM errors}
	\label{sec:Environment-MN}
	\input{20-Environment-MN.tex}

	\section{Mont-Blanc DRAM errors}
	\label{sec:Environment-MB}

\input{25-Environment-MB.tex}

	\section{Timeline analysis}
	\label{sec:TimelineAnalysis}

\input{50-TimelineAnalysis}

	\section{Correlations}
	\label{sec:correlation}

\input{51-Correlations}

 	\section{Periods with the highest neutron counts}
 	\label{sec:ks_tests}

\input{52-KS_tests}

	\section{DRAM error prediction}
	\label{sec:ml}
	\input{55-ML}
	\section{The signal and the noise} 
	\label{sec:signal-noise}
	\input{56-Signal-noise.tex}

	\section{Related Work}
	\label{sec:Related_work}

\input{60-Related_work.tex}

	\section{Conclusions}
	\label{sec:Conclusions}

\input{70-Conclusions.tex}

	\section*{Acknowledgment}
The research was supported by the Spanish Government through the contracts PID2019-107255GB-C21,  PID2023-146511NB-I00 and CEX2021-001148-S, funded by MCIN/AEI/10.13039/501100011033. The work also received funding from the Department of Research and Universities of the Government of Catalonia (Code: 2021 SGR 00807, 2021 SGR 01007, 2021 SGR 01264). Paul Carpenter holds the Ramon y Cajal fellowship RYC2018-025628-I funded by MICIU/AEI/10.13039/501100011033 and "ESF Investing in your future".
Support also came from the European Union’s Horizon Europe research and innovation programme under grant agreement No. 101097224 (REBECCA). Further financial assistance was provided by the Barcelona Zettascale Laboratory (reference REGAGE22e00058408992), which is financed by the Ministry for Digital Transformation and Public Services within the framework of the Resilience and Recovery Facility, and the European Union NextGenerationEU initiative.

We acknowledge the NMDB database (www.nmdb.eu), founded under the European Union's FP7 programme (contract no. 213007), for supplying part of the neutron monitor data. Also, the Space Research Group (SRG-UAH) at the University of Alcala, Spain, for providing the CaLMa neutron monitor data. We further extend our gratitude to Darko Zivanović and Ferad Zyulkyarov for their collaboration and support.


	\bibliographystyle{IEEEtran}
	\bibliography{particles}

\end{document}

%% file: 10-Introduction.tex
\looseness -1 In large-scale compute clusters, main memory is one of the
principal causes of hardware failures~\cite{HP:2016,
Giurgiu:Middleware2017, Schroeder:SIGMETRICS2009, Hwang:ASPLOS2012, Beigi:HPCA2023}.   
These failures are especially costly  
in high-performance computing~(HPC) systems, where a single tightly-coupled
job may execute for days on thousands of nodes. If one of these nodes fails,
the whole job is terminated. It is therefore important to understand memory system reliability, as it is an  
important limit on the ability to scale to larger systems.

Many studies analyze DRAM errors in the field.  
One of the potential causes of DRAM errors are cosmic rays.
Various industrial studies performed between the 1970s and early 1990s~\cite{lage1993soft, o1994effect}
demonstrate that some of these particles have sufficient energy to penetrate Earth's atmosphere down to the sea level, through the ceiling of the multistory buildings and induce DRAM bit-flips~\cite{ziegler1996ibm}. 
Since then, the community has accepted that cosmic rays are a plausible cause of DRAM errors in HPC systems~\cite{Zivanovic:MEMSYS2019, Schroeder:SIGMETRICS2009, Li:USENIX2010, Giurgiu:Middleware2017, Gupta:SC2017, das2021systemic, Levy:SC2018, Du:ICCD2021, li2022correctable}. 

Some studies go a step further and suggest that cosmic rays could explain some aspects of the observed DRAM error behavior, 
such as the higher error rates in the top rack chassis~\cite{Sridharan:SC2013} 
or in the clusters geographically located at higher altitudes~\cite{Schroeder:TDSC2010, Sridharan:SC2013, Sridharan:ASPLOS2015}.  
The studies also suggest that some (but not all) DRAM devices are susceptible to transient faults from cosmic ray strikes, 
and that this susceptibility highly varies between the device manufacturers~\cite{Sridharan:ASPLOS2015}.  
The cosmic rays are also considered a potential cause of failures with no prior symptoms and no relevant information in the system logs~\cite{das2021systemic}.

To the best of our knowledge, only one previous study has analyzed the relationship between DRAM error rates and the cosmic rays intensity. 
El-Sayed and Schroeder~\cite{el2013reading} analyze error logs from HPC clusters at Los Alamos National Lab
and conclude that rates of uncorrected DRAM errors 
are not associated with cosmic rays intensity. 
The authors find this result unexpected 
and provide one possible explanation for it: while increasing cosmic rays might lead to a
higher rates of DRAM corruptions, the types of corruption caused by those events might usually be corrected with the built-in error correction codes. 
This reasoning is aligned with the idea that transient errors typically lead to a single-bit data corruption~\cite{Li:USENIX2010}. 

Our study describes the underlying sources of high-energy cosmic rays, explains which particles should be observed, and points to the publicly-available logs from reliable measurement facilities~(Section~\ref{sec:CosmicRays}). 
We study the DRAM errors of MareNostrum~3 large-scale production system and mid-scale Mont-Blanc prototype, which does not support memory error correction in hardware.   
The systems are located in the same geographical location, in Barcelona~(Spain), 
and the errors are logged in overlapping time-periods. 
The MareNostum~3 logs cover 2000\,billion\,MB-hours of server-class DDR3-1600 DIMMs during which we detected 4.5\,million corrected and 71\,uncorrected DRAM errors~(Section~\ref{sec:Environment-MN}).\footnote{Megabyte-hours (MB-hours) combines both the amount of memory used and the duration for which it is used. Specifically, it quantifies how much memory (in megabytes) is consumed over a specific period of time (in hours).} 
In the Mont-Blanc prototype, we collected 25\,million errors during the 135\,million\,MB-hours production of mobile LPDDR3 memory~(Section~\ref{sec:Environment-MB}). 

We explore a correlation between the cosmic rays intensity and number of corrected, transient corrected and uncorrected DRAM errors~(Section~\ref{sec:TimelineAnalysis}~and~\ref{sec:correlation}). 
We also use statistical tests to study whether the error rates have 
different distributions during periods of elevated cosmic ray activity, 
e.g. above $90^{th}$ or $99^{th}$ neutron count percentile~(Section~\ref{sec:ks_tests}).
Our exploration covers a vast number of tests with numerous error categories, time windows and system scopes. 
We also examine whether machine learning  methods for DRAM error prediction would be enhanced if they 
considered the neutron count measurements (Section~\ref{sec:ml}).  
%
Finally, we apply the presented methodology to verify and question some findings from previous studies 
(Section~\ref{sec:signal-noise}). 
 
We hope that the insights gained in this study and the presented methodology will become a standard for any future analysis of the relationship between cosmic rays and DRAM errors in the field. The analysis can (and should) be also applied to revisit and extend numerous previous studies which use cosmic rays as a hypothetical explanation for some aspects of the observed DRAM error behaviors. 
%
Our study motives, guides and enables a follow-up research on this frequently mentioned, but insufficiently studied phenomenon.

%% file: 15-Background.tex
\subsection{Background}
Cosmic rays are high-energy particles that originate from outer space. 
They normally come from supernovae or other energetic events in the galaxy, and occasionally come from the Sun during periods of high solar activity. 
When cosmic rays enter the Earth's atmosphere, they produce a cascade of secondary particles which can reach the Earth's surface and even penetrate underground. 
The particles hitting the DRAM chips can alternate the cell charge and corrupt the stored data, causing DRAM errors. 

The density of the atmosphere decreases exponentially with altitude, providing less shielding at higher altitudes. This results in a greater flux of secondary particles 
and an increased probability of errors caused by cosmic rays at higher elevations~\cite{ziegler1996terrestrial, ziegler1996ibm, o1994effect}.
A common belief is that cosmic rays are also a plausible cause of DRAM errors in datacenters, especially if they are located on high altitudes. 
This belief, however, is not supported by any large-scale field study. 

\subsection{Data collection}
\label{sec:Cosmic-ray-data-collection}
\looseness -1 From all the cascading elements produced by cosmic rays, neutrons are the most commonly observed at ground level. Their flux can be measured with specialized \textbf{neutron monitors}, ground-based detectors that measure the neutrons that reach the Earth's surface.
Rare solar events that produce very high-energy particles can also generate secondary cascades in the atmosphere, which can be detected by neutron monitors.
%
%
The Neutron Monitor Database~\cite{website:NMDB} provides public access to neutron count logs from 
more than 50 neutron monitors located all over the world, most of them in the northern hemisphere.
 
To quantify the impact of cosmic rays on HPC cluster failure rates, users can correlate the system error logs with the neutron counts from a monitor. The monitor should be selected based on the distance
and the altitude difference with respect to the HPC facility. 
If various nearby neutron monitors are available, the users should compare their measurements to ensure that they are representative of the cosmic ray activity in the region. The exact number of neutron counts reported by these monitors can vary, but their correlation should be high. 

We considered the cosmic ray logs from two neutron monitors. 
The first one is located in Guadalajara~(Spain), 470\,km southwest of Barcelona, situated at an altitude of 708\,meters~\cite{website:CALMA}. 
The second one is in Rome~(Italy), 870~km east of Barcelona and located close to sea level~\cite{website:ROME}, as the MareNostrum supercomputer and the Mont-Blanc prototype. 
During the period of our DRAM error logs, we observed a high correlation coefficient between neutron counts from Guadalajara and Rome. 
%
This correlation, together with the geographic positioning of Guadalajara and Rome neutron monitors, both relatively close to Barcelona but on opposite sides, indicate that the considered neutron monitor measurements are likely representative of the general cosmic rays
in our region. 

Because of proximity, we used the Guadalajara cosmic ray logs.
To ensure that the measurements accurately reflect cosmic ray activity, 
we use data corrected for pressure and instrumental factors. In the Neutron Monitor logs, this is referred to as efficiency correction. 
Corrections for instrumental changes, such as modifications of the building and the measurement equipment, are necessary to maintain data consistency. Also, the atmospheric pressure correction is important because higher atmospheric pressure increases the air mass above the detector, reducing the detected neutron flux. 

Solar cycles are periodic fluctuations in solar activity, lasting approximately 11\,years.
During the period of our logs, we were in a descending phase of a Solar cycle which began in December 2008 and ended in December 2019. 
%
During periods of high solar activity, the solar wind acts as a shield against galactic cosmic rays from sources beyond the Sun and prevent them to reach the Earth~\cite{sierra2018cross}.
A clear anti-correlation has been observed between the activity of the solar cycle and measurements, on the ground and in space, of galactic cosmic rays.
This is visible in the neutron counts, which moderately increase during the observation period (see Section~\ref{ssec:neutron_counts}).

%% file: 20-Environment-MN.tex


\subsection{MareNostrum~3 supercomputer}
\label{sec:SystemConfiguration}

\looseness -1 
The MareNostrum~3 error logs cover a period of more than two years, from October 2014 to November 2016, 
during which we detected 4.5\,million corrected and 71\,uncorrected DRAM errors.\footnote{The log also contains 262 over-temperature DRAM errors, but they are not analyzed because they are unrelated to the cosmic ray activity.}   
The supercomputer was located in Barcelona (Spain), 100\,meters above the sea level,  
in a stone-walls chapel with a ceramic tiles roof. 
At the time, MareNostrum~3 was one of the six Tier-0 (largest) HPC systems in
the Partnership for Advanced Computing in Europe (PRACE)~\cite{website:prace}.
It comprised 3056 compute nodes, each with two eight-core Intel Sandy Bridge-EP
E5-2670 sockets with a 2.6~GHz nominal clock frequency. 
We use the logs from the compute nodes only, excluding the login and test nodes
which are not part of the same monitoring infrastructure and whose failures
do not impact large-scale compute jobs. 

%
%
The MareNostrum~3 compute nodes included more than 25,000 DDR3-1600 DIMMs from all three major manufacturers: 
6694, 5207 and 13,419~DIMMs from \emph{Manufacturer~A},~\emph{B} and~\emph{C}, respectively. The manufacturers have been anonymized to protect the interested parties.  
%
%
MareNostrum~3 DIMMs were built in three different DRAM technologies: $\overline{3x}$\,nm, $\overline{2y}$\,nm and $\overline{2z}$\,nm.  
We are allowed to show only the first of two digits of the nanometer technology and their relative order: 
$\overline{3x}$\,nm $>$ $\overline{2y}$\,nm $>$ $\overline{2z}$\,nm. 
%
During the observation period we collected measurements on more than 2000\,billion\,MB-hours.  
The main workloads executed on MareNostrum~3 were large-scale
scientific HPC applications and the system utilization typically exceeded 95\%.
%

\looseness -1 MareNostrum~3 employed a Chipkill
ECC scheme, which could correct all errors coming from a single x4 device. 
For x8 devices, the ECC can correct up to 4-bit errors coming from the same DRAM chip.  
%
%
The ECC check is performed on each application memory read 
and by a patrol scrubber which periodically traverses the whole physical memory 
and performs an ECC check on each location.    


During production, any DIMM that showed early signs of failure was flagged by a
pre-failure alert and retired by the system administrators. This action was
recorded in the system log together with the date and time.  Over the two-year
period analysed by this paper, 51~DIMMs were retired for this reason.

%
%

\subsection{Data collection} 
\label{sec:DataCollection}

\looseness -1 \textbf{Corrected errors~(CEs)} were logged by a daemon, based on the
{\small\textsf{mcelog}} Linux kernel module~\cite{Kleen:LK2010}, that
periodically extracts information about corrected errors from the CPU
machine-check architecture~(MCA) registers~\cite{Kleen:LK2010}. Each corrected
error was recorded in a log file, which specifies the error timestamp, node
and DIMM id, and the physical location of the error in the DIMM including
rank, bank, row and column.  
The log entry also indicates whether correction
was done during an application memory read or by patrol scrubbing.  

\looseness -1 The daemon accessed the MCA registers with a period of 100\,ms, which was the shortest time interval that caused negligible 
overhead to the production applications.  If more than one error occurred in a
100\,ms time interval, the MCA registers record the number of errors, but
only provide detailed information for one error in the interval.  Our logs
therefore specify the exact total number of corrected errors and provide detailed
error information for a subset of the errors. Increasing the sampling frequency
would increase the number of errors with detailed information, but also
increase the performance overhead of the error logging daemon. Previous studies
perform similar readings of the memory error registers with a period of a few
seconds \cite{Sridharan:SC2012, Sridharan:SC2013, Sridharan:ASPLOS2015} or once
per hour~\cite{Li:USENIX2010}.

\looseness -1 Our analysis covers both, \textbf{transient} and \textbf{non-transient} CEs. Transient CEs 
are frequently considered to be caused by temporary environmental factors, such as cosmic ray 
strikes~\cite{Li:USENIX2010, Hwang:ASPLOS2012, Sridharan:SC2012, Sridharan:ASPLOS2015}. 
We consider an error to be transient if, during the entire observation period, the error occurred only once in a given cell and no other errors occurred in the same row or column~\cite{Hwang:ASPLOS2012}.

\textbf{Uncorrected errors~(UEs)} were logged by IBM
firmware~\cite{IBM:iDataPlex:2014}, which is part of the MareNostrum~3
monitoring software.  For each uncorrected error, the log specifies the DIMM
that failed and the cause of the error, i.e. whether it occurred during
an application memory read or patrol scrubbing. 
Finally, the log records UE warnings
generated when memory modules are throttled to prevent an over-temperature condition
and when the correctable ECC logging limit has been reached. 
   
%

%% file: 25-Environment-MB.tex
\subsection{Mont-Blanc prototype}

The prototype was located in Barcelona (Spain), in the same area and altitude as MareNostrum, 100\,meters above the sea level. 
The system  was deployed in a brick building with a concrete roof.
The Mont-Blanc prototype comprised 1080 nodes based on the Samsung Exynos 5250 mobile system-on-chip~(SoC)~\cite{Rajovic:SC2016}.  
Each SoC integrated two 1.7\,GHz Arm Cortex-A15 cores, one on-chip Mali-T604 GPU and 4\,GB of LPDDR3-1600. 
All memory devices are from the same manufacturer. 
%
The objective of the Mont-Blanc prototype was to explore the suitability of high-performance computing on hardware initially targeting mobile devices. 
Therefore, the deployed LPDDR3 main memory included no error detection or correction mechanisms. 

\subsection{Data collection} 
\label{sec:MB-data-collection}

Mont-Blanc error logging lasted for over a year, February 2015 to February 2016, which is within the time-frame of the MareNostrum logs. It monitored over 4.2\,million node-hours, 135\,million\,MB-hours of LPDDR3-1600, and logging over 
25\,million memory errors~\cite{bautista2016unprotected}.     
DRAM errors were logged with a software daemon running an infinite loop that traverses the whole available system memory. 
In each node, the daemon allocated approximately 3\,GB of main memory, while the rest was dedicated for the OS and other system software.  
The daemon was first writing a specific value into each memory location, and then, in the next loop iteration, was checking whether the memory still contained the correct data. If this was not the case, the tool recorded the memory error location and the number of corrupted bits. 
Since the daemon memory utilization was significant, it was executed only when the nodes were not running any user job. 
For this reason, the total amount of memory scanned depended on the prototype utilization, which varied in time. 
To account for this, we normalize the number of DRAM errors per amount of scanned memory in that specific time. 
%
The Mont-Blanc error-logging framework was set by Bautista et al.~\cite{bautista2016unprotected}. 
Our analysis is based on the data reported in the original study.

%% file: 50-TimelineAnalysis.tex
We first examine the timelines of the DRAM error logs and neutron counts to understand the general trends and overall behavior. This analysis helps to put into context the correlation analysis of Section~\ref{sec:correlation}.

\subsection{Neutron counts}
\label{ssec:neutron_counts}


Figure~\ref{fig:calma_logs} shows neutron counts per second recorded by the Guadalajara neutron monitor 
from October~2014 to November~2016, the period corresponding to the DRAM error logs of MareNostrum~3.   
The overall neutron count range 
in this period is representative of the complete operational neutron monitor data spanning from 2012 to 2024.  
We plot the neutron counts per second averaged by hour and by month. 
The hourly data is the finest granularity that provides a readable plot. 
We also see some variation within the cycle, and some peak behavior, e.g. negative peaks in January and July~2015, 
and a positive peak in November~2016.  
Finally, we also see a general trend of increasing monitor counts over time, especially after August~2015. 
This increasing trend is confirmed with the trend-line showing monthly neutron averages, 
and it may be attributed to decreasing solar activity in this period, as discussed in Section~\ref{sec:Cosmic-ray-data-collection}.

\subsection{DRAM errors}

\begin{figure}[t!]
	\centering
	\includegraphics[width=\columnwidth]{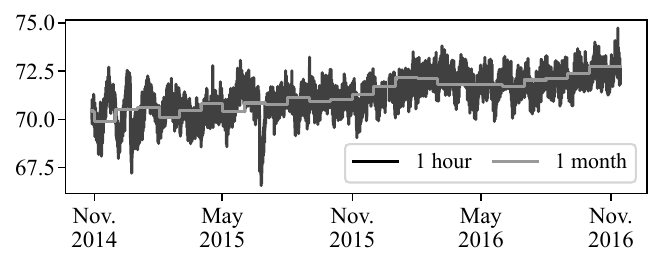}
	\caption{Neutron counts per second. Hour and month averages.}
	\label{fig:calma_logs}
\end{figure}
\begin{figure}[h!]
	\centering
	\begin{subfigure}[b]{\columnwidth}
		\centering
		\includegraphics[width=\columnwidth]{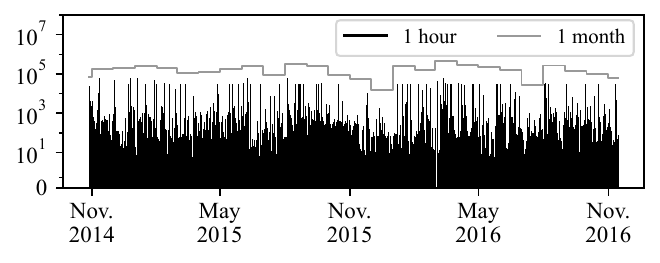}
		\caption{MareNostrum~3: Total corrected errors per hour and per month. }
		\label{fig:ces_logs}
	\end{subfigure}
	
	%
	\begin{subfigure}[b]{\columnwidth}
		\centering
		\includegraphics[width=\columnwidth]{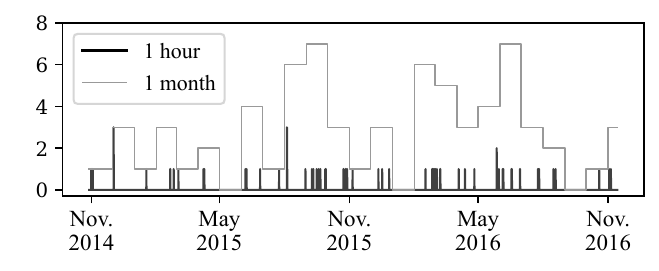}
		\caption{MareNostrum~3: Total uncorrected errors per hour and per month.}
		\label{fig:ues_logs}
	\end{subfigure}
	
	
	\begin{subfigure}[b]{\columnwidth}
		\centering
		\includegraphics[width=\columnwidth]{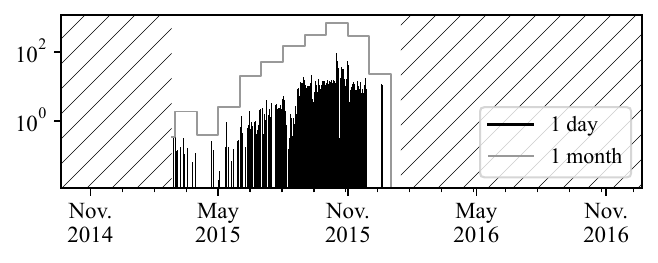}
		\caption{Mont-Blanc: Total errors per day and per month.}
		\label{fig:montblanc_logs}
	\end{subfigure}
	
	\caption{DRAM errors over time. MareNostrum~3 production supercomputer and Mont-Blanc prototype.}
	\label{fig:dram_logs}
	
\end{figure}

Figure~\ref{fig:ces_logs} shows the \textbf{MareNostrum corrected errors~(CEs)} hourly and monthly occurrences 
between October~2014 and November~2016. 
On most hours there are up to 100 corrected errors, but on a few hours the number of errors is large, up to about 10,000. 
The results suggest that we should further explore a potential alignment between the neutron count and error peaks.
%
Comparison of the neutrons and CEs monthly step-lines, Figures~\ref{fig:calma_logs} and~\ref{fig:ces_logs}, 
detects no obvious correlation between these two variables. 
We repeat this analysis focused only on the transient CEs (not depicted in Figure~\ref{fig:dram_logs}), 
and we reach the same conclusions. 

Figure~\ref{fig:ues_logs} shows the number of \textbf{uncorrected MareNostrum DRAM errors~(UEs)} per hour and per month, across all DIMMs. %
In most hours we detect no errors, on a few hours we detect
one error, and very occasionally we detect two or three errors. %
In total during the observation period of 25 months
we detect only 71 errors.
%
Regarding the monthly values, we detect somewhat higher UE rates in the period June~2015 to June~2016, which does correspond 
to above-average neutron counts. However, we also detect some of the lowest UE numbers  
in the last months of the study, which is the period of the highest cosmic rays intensity. 
Overall, the visual comparison of the neutron and UE trend-lines detects no obvious correlation.

Figure~\ref{fig:montblanc_logs} depicts the number of errors per day and per month for the Mont-Blanc system, 
normalized to the amount of scanned main memory.
The timescale ($x$-axis) is aligned with the neutron counts measurements in Figure~\ref{fig:calma_logs}
and the hatched section of the chart indicates the time period for which we do not have Mont-Blanc error logs. 
From May to November 2015, we see a significant monotonic increment of the error rates, 
well correlated with the neutron counts.   
However, the correlation is low at the beginning and the end of the Mont-Blanc logs. 
This is especially pronounced in the last three months, November~2015 to February~2016, in which the
neutron counts continue to increase, while the error rates drop by an order of magnitude.  


%% file: 51-Correlations.tex

Next, we further explore the correlation between the neutron counts and DRAM error
rates.

\subsection{Methodology}
\label{ssec:corr_methodology}

First, we use a visual correlation between the neutron counts and the total number of DRAM errors detected in the same time period.  
Second, we apply the Kendall test~\cite{kendall1938new} to explore correlation for a large number of error 
categories.\footnote{Kendall's test assesses the strength and direction of the association between two variables
by measuring the ordinal association between them.
We also considered Pearson correlation and Spearman’s rank correlation. 
Pearson correlation is disregarded because it requires normal data distribution and a linear relationship between the two variables.  
As Kendall's coefficient, Spearman’s correlation is a rank-based non-parametric test, so it can be applied to our analysis. 
We selected the Kendall correlation measure because it is more robust than Spearman’s rank correlation~\cite{kendall-vs-spearman}.}   
We perform statistical tests on all combinations of the error categories, time windows, and system scopes listed in Table~\ref{tab:test_combinations} (MareNostrum~3) and Table~\ref{tab:montblanc_test_combinations} (Mont-Blanc). 
For MareNostrum supercomputer, this leads to a vast number of correlation tests: 1,764,096 for uncorrected and 10,584,576 for
 corrected errors.\footnote{For example, for uncorrected errors, there are four time-windows, four manufacturer categories, 
three error types, and four technologies for a total of 192 combinations.  
\looseness -1 The main increase in tests comes from 9188 different system scopes. 
The tests are applied to the whole system, but also individually for each of the 37 racks, 3050 nodes, and 6100 sockets ($1+37+3050+6100 = 9188$).} 
In Mont-Blanc error logs contain much less information, leading to only 21 error categories.

\begin{table}[th!] 
	\caption{MareNostrum~3 error categories, time-windows and system scopes.}
	\begin{center}
		\small
		\begin{tabular}{@{}lp{0.55\columnwidth}@{}}
			\toprule
			\multicolumn{2}{@{\hskip 0pt}l}{\textbf{MareNostrum uncorrected and corrected errors }} \\
			\midrule
			Manufacturers & $A$, $B$, $C$, All \\
			DIMM technology & $\overline{3x}$\,nm, $\overline{2y}$\,nm, $\overline{2z}$\,nm, All \\
			Time window & Hour, day, week, month \\
			System scope & Whole system, 37 racks, 3050 nodes, 6100\,sockets \\
			
			\midrule
			\multicolumn{2}{@{\hskip 0pt}l}{\textbf{MareNostrum corrected errors}} \\
			\midrule
			Transient \emph{vs.} Non-transient & Transient, Non-transient, All \\
			Error type & Memory read, Patrol scrub, All \\
			Single/multi-cell & Single-cell, Multi-cell, All \\
			CEs \emph{vs.} DIMMs & Total number of CEs, Number of DIMMs with at least one CE \\
			
			\midrule
			\multicolumn{2}{@{\hskip 0pt}l}{\textbf{MareNostrum uncorrected errors}} \\
			\midrule
			Error types & Uncorrected ECC, Scrub failed, All \\
			\bottomrule
		\end{tabular}
		\label{tab:test_combinations}
	\end{center}
\end{table}


\begin{table}[th!] 
	\caption{Mont-Blanc error categories and time-windows.}
	\begin{center}
		\small
		\begin{tabular}{@{}lp{0.65\columnwidth}@{}}
			\toprule
			n-bit error & 1, 2, 3, 4, 5, 6+ bits, All \\
			Time window & Day, week, month \\
			\bottomrule
		\end{tabular}
		\label{tab:montblanc_test_combinations}
	\end{center}
\end{table}


%
\looseness -1 For each experiment, the Kendall's test returns a correlation coefficient between -1 (strong inverse correlation) and +1 (strong direct correlation). 
The test also detects cases in which the correlation cannot be tested.  
If all values of one variable are identical, e.g. zero detected DRAM errors, it is impossible to rank them meaningfully, 
and there is no basis to determine concordance or discordance with the cosmic rays intensity.
Finally, the Kendall's test returns a \textit{p}-value that quantifies the correlation significance. 
A large number of applied tests increases the probability 
that some tests incorrectly suggest correlation significance (i.e. low \textit{p}-values). 
To minimize this likelihood 
we apply the Benjamini--Yekutieli false discovery rate method~\cite{benjamini2001control}. 

\subsection{Visual correlation}

\begin{figure}[t!]
	\centering

	\begin{subfigure}[b]{\columnwidth}
		\centering
		\includegraphics[width=\columnwidth]{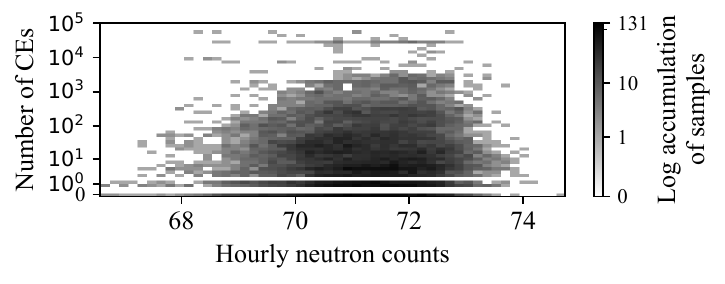}
		\caption{MareNostrum: Corrected errors}
		\label{fig:ces_heatmap}
	\end{subfigure}

	\begin{subfigure}[b]{\columnwidth}
		\centering
		\includegraphics[width=\columnwidth]{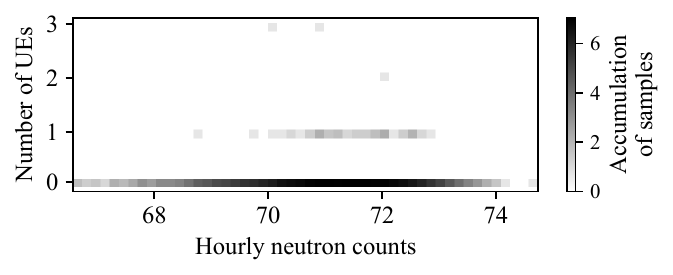}
		\caption{MareNostrum: Uncorrected errors}
		\label{fig:ues_heatmap}
	\end{subfigure}
	
	\begin{subfigure}[b]{\columnwidth}
		\centering
		\includegraphics[width=\columnwidth]{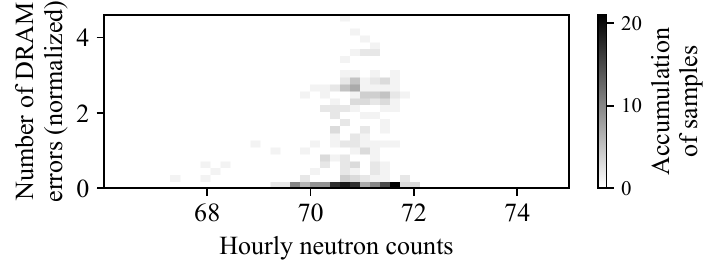}
		\caption{Mont-Blanc errors (normalized)}
		\label{fig:montblanc_heatmap}
	\end{subfigure}

	\caption{Neutron counts ($x$-axis) and the number of DRAM errors in the corresponding time period ($y$-axis). Bare-eye analysis detects no correlations. } 
	\label{fig:error_heatmap}
\end{figure}

Figure~\ref{fig:error_heatmap} shows averaged neutron counts per second ($x$-axis) and the number of memory errors recorded in the corresponding time period ($y$-axis). As in the rest of the paper, the Mont-Blanc error counts are normalized with the amount of scanned memory. 
The figures plot the data for the whole system, 3056 MareNostrum~3 nodes and 1080 Mont-Blanc nodes. 
We use a gray heat-map to show the number of observations with a given error and neutron count.  
To cover a large range of MareNostrum CEs per hour, the $y$-axis and the heatmap in Figure~\ref{fig:ces_heatmap} use a log scale.

Overall, we find no correlation that can be observed with the naked eye. 
We repeat the experiments for MareNostrum transient corrected errors, and we reach the same conclusion. 
All three charts show a higher density between 70 and 72 neutron counts per second 
because this is the most frequent neutron count range during the time of the study.

\subsection{Statistical tests}
\label{sec:Correlation-statistics}

\begin{figure}[t!]
	\centering
	
	\begin{subfigure}[b]{\columnwidth}
		\centering
		\includegraphics[width=\columnwidth]{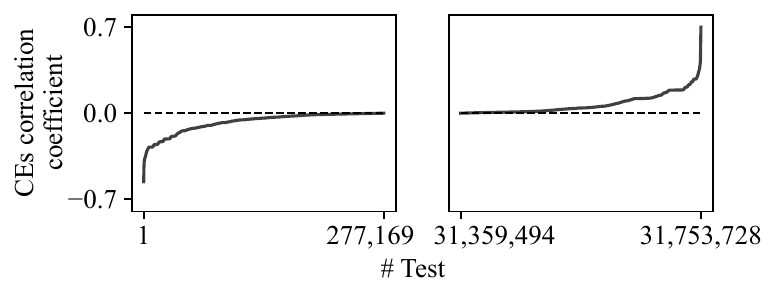}
		\caption{MareNostrum: Corrected errors}
		\label{fig:ces_kendall_corrs}
	\end{subfigure}
	
	\begin{subfigure}[b]{\columnwidth}
		\centering
		\includegraphics[width=\columnwidth]{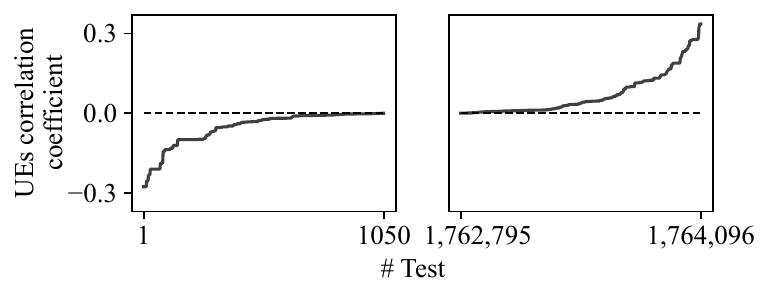}
		\caption{MareNostrum: Uncorrected errors}
		\label{fig:ues_kendall_corrs}
	\end{subfigure}
	
	\begin{subfigure}[b]{\columnwidth}
		\centering
		\includegraphics[width=\columnwidth, trim=0 0 -0.6cm 0]{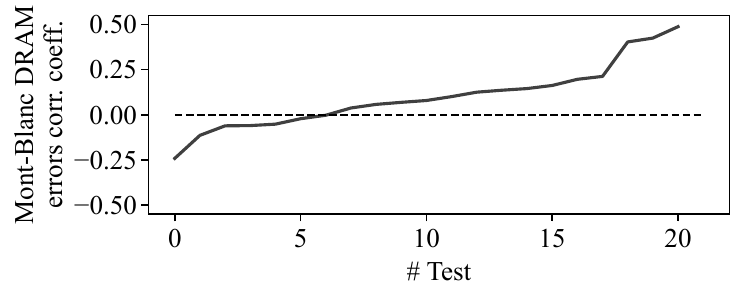}
		\caption{Mont-Blanc errors}
		\label{fig:montblanc_kendall_corrs}
	\end{subfigure}
	
	\caption{Most of the correlation tests report low coefficients.}
	\label{fig:kendall_corrs}
\end{figure}

\looseness -1 In \textbf{MareNostrum} analysis, numerous error categories, time windows and system scopes lead to a vast overall exploration space. 
However, the nature of the DRAM errors and some practical limitations, 
hugely reduce the number of practical correlation tests: 
from 1,764,096 to 2352 for uncorrected, and 
from 31,753,728 to 671,404 for corrected errors.
There are three main reasons for this reduction. 
First, most of the nodes experienced no DRAM errors: 62\% experienced no CEs,  
and 99\% no UEs.  
For these nodes, there is no basis to determine DRAM error concordance or discordance with the neutron counts.  
Also, not all possible categorical combinations are present in practice.  
For example, each MareNostrum node comprises DIMMs from the same manufacturer and technology. 
If a given node experiences DRAM errors, it can be tested only for this specific manufacturer--technology pair,  
not for all possible combinations of these two parameters. 
Finally, many of the error log inputs are incomplete, with some fields that are omitted.

Figure~\ref{fig:kendall_corrs} shows the \textbf{Kendall's correlation coefficient} for all DRAM error categories, time windows and system scopes. 
The coefficients are sorted in ascending order.
To increase the visibility in MareNostum experiments, we plot distinct charts for negative and positive ($\geq0$) correlation coefficients, 
while the gap between them indicates the tests that could not be deployed in practice.
The corrected error coefficients range are low, with only a handful of experiments that exceed a value of $0.3$, see Figure~\ref{fig:ces_kendall_corrs}. 
For the uncorrected errors, depicted in Figure~\ref{fig:ues_kendall_corrs}, the coefficients are even lower. 
Also, in both charts we observe a roughly equal proportion of the positive correlations, and the unexpected negative ones.  

For the \textbf{Mont-Blanc prototype}, we test 21 correlations, for three time-windows and seven error types. 
Their correlation coefficients range from $-0.24$ to $+0.49$. 

\begin{figure}[t!]
	\centering
	
	\begin{subfigure}[b]{\columnwidth}
		\centering
		\includegraphics[width=\columnwidth]{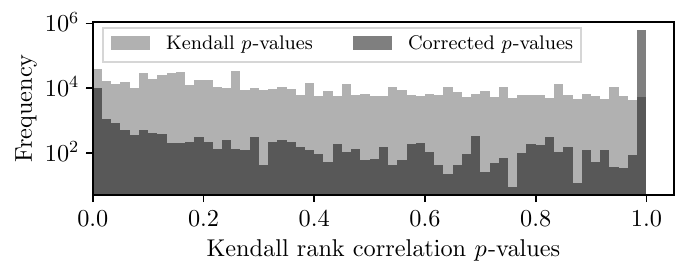}
		\caption{MareNostrum: Corrected errors}
		\label{fig:ces_correlation_pvals}
	\end{subfigure}
	
	\begin{subfigure}[b]{\columnwidth}
		\centering
		\includegraphics[width=\columnwidth]{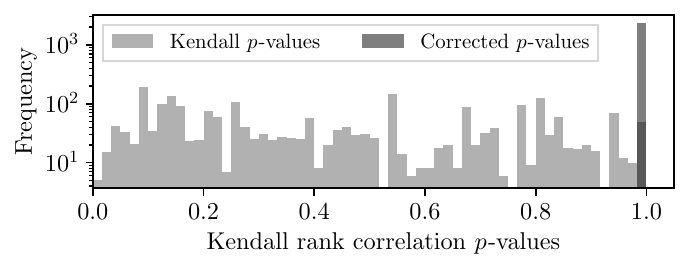}
		\caption{MareNostrum: Uncorrected errors}
		\label{fig:ues_correlation_pvals}
	\end{subfigure}
	
	\begin{subfigure}[b]{\columnwidth}
		\centering
		\includegraphics[width=\columnwidth]{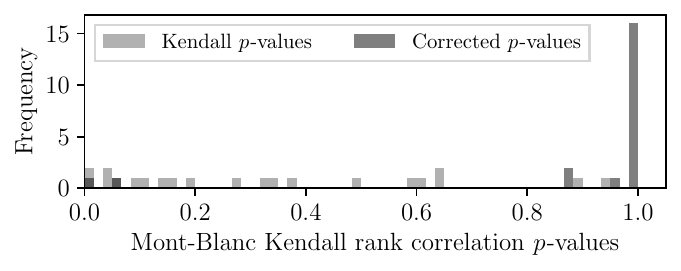}
		\caption{Mont-Blanc errors}
		\label{fig:montblanc_correlation_pvals}
	\end{subfigure}
	
	\caption{After the false discovery rate correction, $p\text{-value}=1$ in a vast majority of correlation tests.}
	\label{fig:correlation_pvals}
\end{figure}

To consider the \textbf{significance} of the Kendall's correlation coefficients, we analyze the corresponding \textbf{\textit{p}-values}. 
Figure~\ref{fig:correlation_pvals} depicts the histogram of the \textit{p}-values provided directly by the Kendall's test 
and after the Benjamini--Yekutieli false discovery rate correction~\cite{benjamini2001control}.

In \textbf{MareNostrum} tests, after the correction, $p\text{-value}=1$ in a vast majority of the CE tests and all UE tests.  
As the final step of this analysis, we explore the tests
that have statistically significant (\textit{p}-value below 0.05) and moderate or high correlation coefficient.
We detect no such tests for transient corrected errors. 
This is surprising because cosmic ray strikes are frequently considered a plausible cause of these errors~\cite{Li:USENIX2010, Hwang:ASPLOS2012, Sridharan:SC2012, Sridharan:ASPLOS2015}.  
%
%
When we analyze the whole CE sample, we find no statistically significant tests with inverse correlations below $-0.5$, 
and only 40 tests with positive correlations above $+0.5$.  
All of them correspond to the 1\,month time window in the 25-months logging period, i.e. the sample of only 25 observations. 
Out of the 40 tests, 36 are from three DIMMs located in two nodes. 
All three DIMMs experience higher error rates in the months of the logging period, coinciding with higher neutron counts. 
However, the recorded errors are repeated,  up to thousands of times, on a small number of distinct physical locations (memory cells). 
Therefore, it is highly unlikely that their underlying cause is an external factor, such as cosmic rays.  

After the false discovery rate correction, only four \textbf{Mont-Blanc} tests have \textit{p}-values below 0.05. 
All of them correspond to correlation coefficients of 0.13 and 0.14, indicating a low correlation with cosmic rays.

The correlation tests consider the overall trend across all data points
and may not reveal a significant association if DRAM errors are more likely to occur only above a certain neutron count threshold. 
To address this limitation, in the next section we explore the error rates during periods of elevated cosmic ray activity.

%% file: 52-KS_tests.tex

\subsection{Methodology}
\label{ssec:ks_tests_methodology}

To explore whether DRAM errors are more affected by cosmic rays above a certain neutron count threshold, 
we partition the DRAM errors into two samples: those observed during high neutron-count periods (above a certain threshold) 
and the remaining errors. 
Then we use the Kolmogorov--Smirnov~(KS) test~\cite{berger2014kolmogorov} to determine whether two DRAM error samples come from the same distribution. 
Since there is no predefined neutron count threshold which is known to impact DRAM errors,  
we consider the $90^{th}$, $95^{th}$, $99^{th}$, and $99.9^{th}$ neutron count percentiles. 
As in the previous section, we apply the test to all combinations of the error categories, time windows and system scopes listed in Table~\ref{tab:test_combinations}~and~\ref{tab:montblanc_test_combinations}, and then use the Benjamini--Yekutieli \textit{p}-value correction.

\subsection{Results}

\begin{figure}[t!]
	\centering

	\begin{subfigure}[b]{\columnwidth}
		\centering
		\includegraphics[width=\columnwidth]{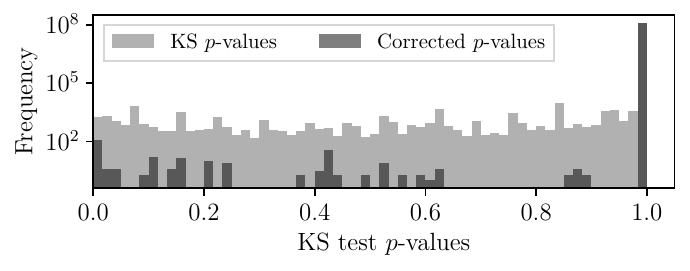}
		\caption{MareNostrum: Corrected errors}
		\label{fig:ces_ks_pvals}
	\end{subfigure}

	\begin{subfigure}[b]{\columnwidth}
		\centering
		\includegraphics[width=\columnwidth]{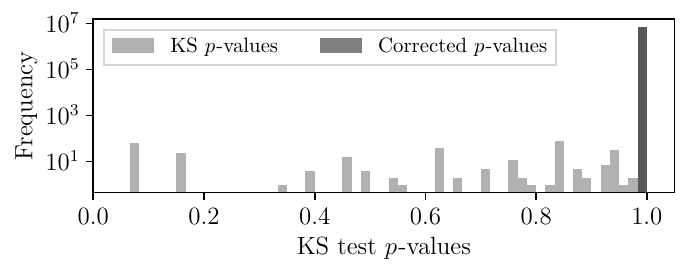}
		\caption{MareNostrum: Uncorrected errors}
		\label{fig:ues_ks_pvals}
	\end{subfigure}

	\begin{subfigure}[b]{\columnwidth}
		\centering
		\includegraphics[width=\columnwidth]{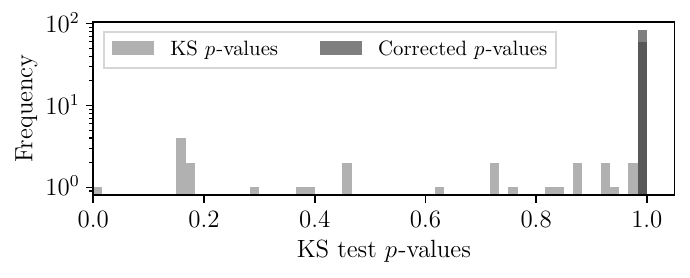}
		\caption{Mont-Blanc errors}
		\label{fig:montblanc_ks_pvals}
	\end{subfigure}
	
	\caption{After the false discovery rate correction, $p\text{-value}=1$ in a vast majority of correlation tests.}
	\label{fig:ks_pvals}
\end{figure}

Figure~\ref{fig:ks_pvals} shows the histogram of the \textit{p}-values provided directly by the Kolmogorov--Smirnov test 
and after the false discovery rate correction.  
After the correction, $p\text{-value}=1$ in all Mont-Blanc and MareNostrum uncorrected error tests. 
Out of 127,014,912 MareNostrum corrected error tests, we detect
40 statistically significant instances (\textit{p}-values below 0.05) that exhibit a surprising behavior: lower error rates during the higher cosmic ray activity.  
The expected behavior, higher error rates during periods with neutron counts above $90^{th}$ and $95^{th}$ percentile, 
is detected in 184~tests.  
Most of these 184~tests are triggered by only five DIMMs located in three nodes. 
These DIMMs and the nodes are different from the ones that had a statistically significant correlation with the neutron counts
(see Section~\ref{sec:Correlation-statistics}). 
However, they show a similar behavior: they experience higher error rates in the months
with the highest cosmic rays activity, but the errors are periodically repeated on a small number of distinct physical locations.  
Therefore, it is highly unlikely that their underlying cause are cosmic rays.  
  
%% file: 55-ML.tex
We also explore whether machine learning methods for DRAM error prediction would be enhanced by
considering the neutron count measurements. 

%

\subsection{Methodology}
\looseness=-1 We start from random forest model used by Boixaderas et al.~\cite{boixaderas2020cost}.
The model predicts \textbf{DRAM uncorrected errors~(UEs)} based on system metrics and error log information from each DIMM: 
number of CEs, UEs and UE warnings, number of affected ranks, banks, rows and columns, etc. 
The model checks the system logs every minute and then predicts whether a UE will occur in a DIMM within the next day. 
The evaluation is based on a cost--benefit analysis, which compares the system resources needed for training, failure prediction and failure mitigation against the saved compute time due to successful failure prediction and mitigation.

We also use the same random forest model, set of features and methodology to predict the forthcoming \textbf{corrected errors~(CEs)}. 
To account for the higher frequency of CEs (compared to UEs) we reduce the prediction window to one hour. 
The objective of our CE predictions is not to improve the HPC system utilization,\footnote{The CEs do not lead to HPC job failures and lost compute time~\cite{boixaderas2020cost}.}  
but to investigate whether the neutron counts data enhance the model's predictive accuracy. 
If this were the case, it would indicate that the deployed random forest model is capable of detecting subtle relationships between cosmic ray intensity and CE rates;
a relationship not captured by the statistical methods analyzed in the previous sections. 
The CE evaluation is based on the Area Under the Curve~(AUC), 
a standard performance metric used to evaluate the quality of a binary classification model~\cite{bradley1997use}.
The metric considers the model's true positive and false positive rates across various decision threshold settings. 

To explore the influence of cosmic rays, we extend the feature set of the original random forest model with the \textbf{neutron counts} data. 
We include the average, standard deviation, and percentage variation of neutron counts over 1\,hour, 5\,hours, 10\,hours, 1\,day, 1\,week, and 1\,month. 
We split the feature data into 60\% for training, 20\% for validation and 20\% for testing. After training and selecting the best model through hyperparameter tuning, we retrain it using both the training and validation sets before evaluating its final performance on the test set. 
To determine the significance of neutron count features, we train and evaluate the primary and reference models.  
The primary models use the \textbf{actual neutron counts} information. 
The reference models use the same data, but with \textbf{permuted neutron count} features. 
This is a standard procedure for assessing feature importance because it maintains the distribution of feature values
and ensures the consistency between primary and reference models~\cite{altmann2010permutation}. 
%
We analyze the predictive capabilities of the models for corrected 
and uncorrected errors, and use the Gini importance~\cite{Breiman:RandomForests} 
to analyze the significance of the neutron count features. 

%


\subsection{Results}

\begin{figure}[t!]
	\centering
	
	\begin{subfigure}[b]{\columnwidth}
		\centering
		\includegraphics[width=\columnwidth]{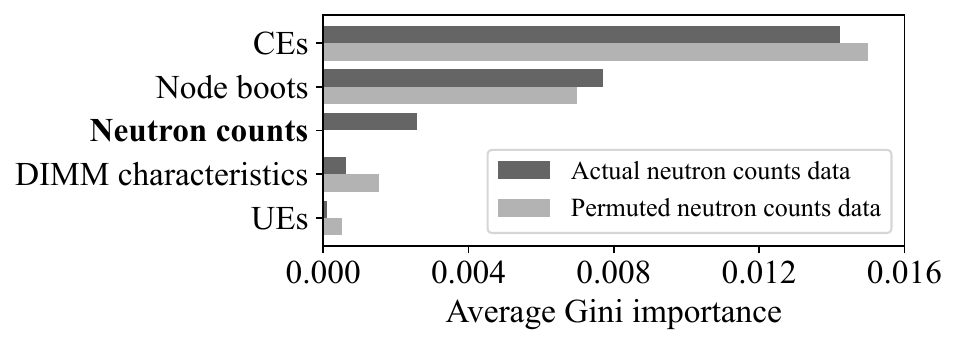}
		\caption{MareNostrum: Corrected errors}
		\label{fig:ces_ft_importances}
	\end{subfigure}
	
	\begin{subfigure}[b]{\columnwidth}
		\centering
		\includegraphics[width=\columnwidth]{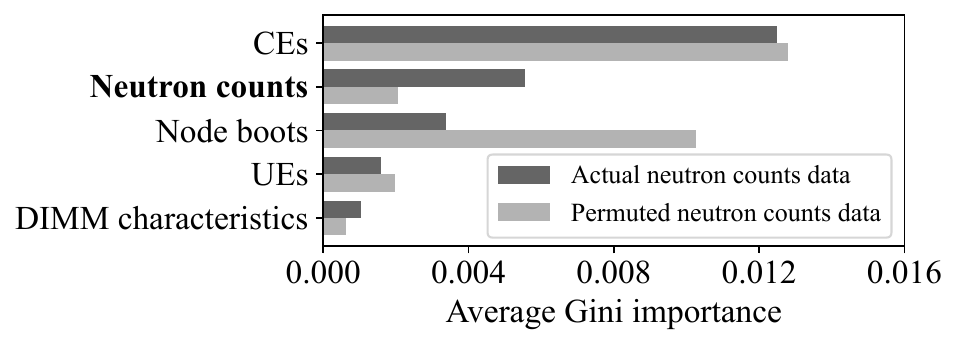}
		\caption{MareNostrum: Uncorrected errors}
		\label{fig:ues_ft_importances}
	\end{subfigure}
	
	\caption{Gini importance indicates which features are most influential in the model's predictions.}
	\label{fig:ft_importances}
\end{figure}

The models that include the actual neutron counts show practically the same prediction performance as the reference models 
with permuted neutron count features. 
The primary and reference CE prediction models achieve an AUC of 0.92 and 0.93, respectively.   
The cost--benefit evaluation of their UE predictions suggest savings of 7451 and 7449~node--hours.\footnote{
We use the same HPC job logs and cost--benefit calculation as the original study of Boixaderas et al.~\cite{boixaderas2020cost}. 
The only difference is that we train the model with the majority of the logging data, and test with UEs from the final 20\% of the logs. The original study uses time series cross-validation across the entire dataset, starting with smaller datasets at the beginning of the logs and incrementally increasing their size. For this reason, we report a different number of saved node--hours.}
%


Figure~\ref{fig:ft_importances} shows the importance of different feature categories, as assigned by the prediction models. 
Our analysis focuses on the neutron counts group (in bold). 
In the CE prediction model with actual neutron counts, this feature category is the third most important one (Figure~\ref{fig:ces_ft_importances}). 
The reference model appropriately assigns zero Gini importance to these permuted features.
The UE models assign even higher importance to the neutron count features (Figure~\ref{fig:ues_ft_importances}). 
In the primary model, this is the second most important category, exceeded only by the CE features.   
Even the reference model puts significant importance to these features.  
This unexpected model behavior could be related to the instability of the UE prediction model. 
This is because a small number of uncorrected errors leads to a significant class imbalance that requires considerable undersampling~{\cite{boixaderas2020cost}, reducing the available information during training and leading to increased model instability.

Overall, CE and UE prediction models using actual neutron counts assign some importance to these features, but they do not perform better than the reference models in which the neutron counts are permuted. 
The assigned importance could be attributed to the inherent randomness of the models, where the neutron counts may occasionally appear significant in random subsets of the data. This can be observed in Figure~\ref{fig:ues_ft_importances}, where even randomly permuted neutron counts are given certain importance. 
Another possibility could be that neutron counts are correlated with another feature, causing the importance to shift to that feature when neutron counts are permuted. However, we discard this possibility by performing correlation tests between each neutron count feature and every other feature, confirming that no such correlation exists.
Finally, it is also possible that neutron counts are only important in a very limited set of circumstances, making their impact imperceptible when evaluating the overall model performance.

%% file: 56-Signal-noise.tex
%

In this section we demonstrate why the analysis and methodology we advocate are so important, 
and what can happen if they are not followed. 
 
\begin{figure}[t!]
	\centering
	\begin{subfigure}[b]{\columnwidth}
		\centering
		\includegraphics[width=\columnwidth]{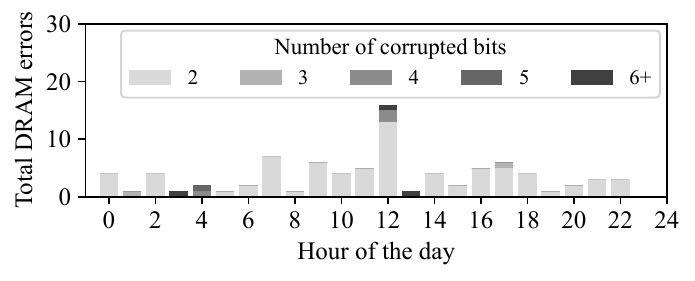}
		\vspace{-0.6cm}
		\caption{Mont-Blanc errors. Reproduced from the original study~\cite{bautista2016unprotected}.}
		\label{fig:montblanc_hour}
	\end{subfigure}

	\centering
	\begin{subfigure}[b]{\columnwidth}
		\centering
		\includegraphics[width=\columnwidth]{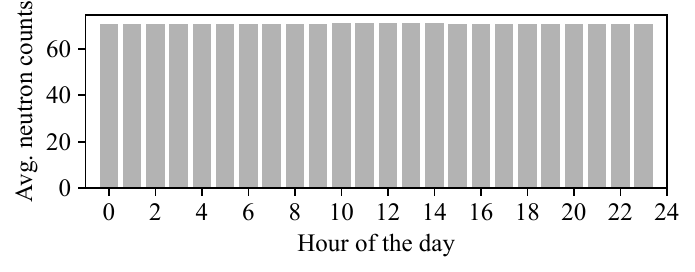}
		\vspace{-0.6cm}
		\caption{Cosmic rays}
		\label{fig:neutrons_hour}
	\end{subfigure}

	\centering
	\begin{subfigure}[b]{\columnwidth}
		\centering
		\includegraphics[width=\columnwidth]{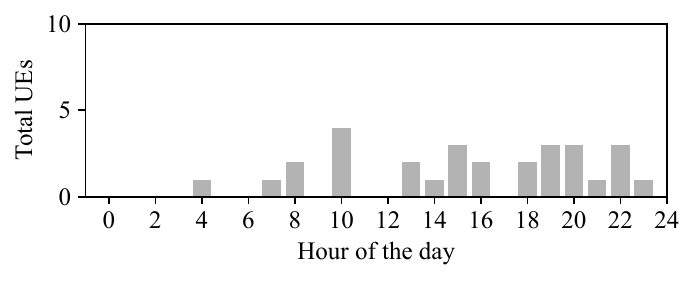}
		\vspace{-0.6cm}
		\caption{MareNostrum: Uncorrected errors}
		\label{fig:ues_hour}
	\end{subfigure}
	
	\begin{subfigure}[b]{\columnwidth}
		\centering
		\includegraphics[width=\columnwidth]{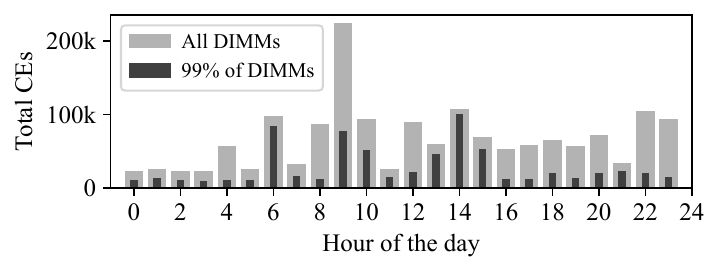}
		\vspace{-0.6cm}
		\caption{MareNostrum: Corrected errors}
		\label{fig:ces_hour}
	\end{subfigure}

	\begin{subfigure}[b]{\columnwidth}
		\centering
		\includegraphics[width=\columnwidth]{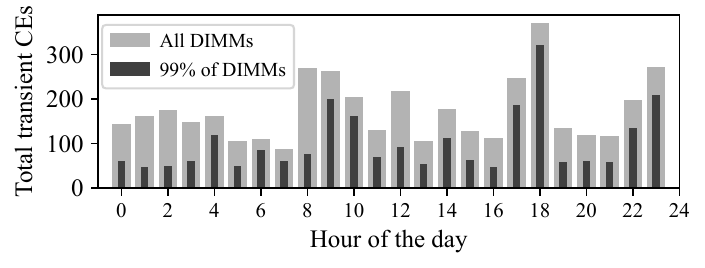}
		\vspace{-0.6cm}
		\caption{MareNostrum: Transient corrected errors}
		\label{fig:ces_transient_hour}
	\end{subfigure}
	
	\caption{The neutron counts do not vary between hours. DRAM error variations cannot be explained with cosmic rays intensity.}
	\vspace{-0.8cm}
	\label{fig:errors_hour}
\end{figure}

Previous study on the Mont-Blanc DRAM errors \emph{``presents evidence which suggests that multi-bit errors are more likely to occur
during day time with a high peak when the Sun is at the highest point in the sky''}~\cite{bautista2016unprotected}. 
The authors therefore \emph{``suggest that multi-bit memory errors are mostly caused by cosmic rays''}. 
These conclusions are based on the results presented in Figure~\ref{fig:montblanc_hour}, which is reproduced from the original paper. 
The figure plots the total number of multi-bit memory corruptions at a given hour of the day ($x$-axis), 
during the observation period from February 2015 to February 2016.
 
\looseness -1 Figure~\ref{fig:neutrons_hour} plots the neutron counts recorded by the Guadalajara neutron monitor in the same period. 
Unsurprisingly, we detect no variation in cosmic ray intensity depending on the time of the day. 
We get the same results for the complete operational neutron monitor data spanning from 2012 to 2024. 
This is simply because the position of the Sun on the sky does not impact the rates of cosmic rays that could cause DRAM errors. 

For comparison, we also plot the hour-of-the-day for MareNostrum errors between February 2015 and February 2016. 
In this period, we detected 29 uncorrected errors (Figure~\ref{fig:ues_hour}).  
%
We also plot the charts for all corrected errors (Figure~\ref{fig:ces_hour}) and the transient ones (Figure~\ref{fig:ces_transient_hour}). 
For all CEs, we detect a clear peak at 9\,h, while the largest number of transient CEs is detected at 18\,h. 
To illustrate how a small number of DIMMs can change the general trend of the data, we remove 1\% of the DIMMs with the highest number of errors.  These results are shown with darker narrow bars in Figures~\ref{fig:ces_hour} and~\ref{fig:ces_transient_hour}. 
We still detect some variation between the time of the day, but some of the highest error rates are now recorded in different hours.  
We repeated the same analysis with our full logs covering the period October 2014 to November 2016. These results are not displayed, but the conclusions are the same.
%

%
%
%

\looseness -1 Our results show that DRAM error rates can be volatile even after long periods of logging on large-scale systems~\cite{Zivanovic:MEMSYS2019}. 
Therefore, although in many research areas bare-eye analysis of the charts might be sufficient, 
in this particular domain, it will almost certainly lead to unreliable results and incorrect conclusions. 
The best way to overcome this problem is to use formal statistical methods. 
The methods that we advocate and deploy are simple to understand and implement, reliable and widely accepted in the statistical community. 
In addition to the extensive methodology, we describe the underlying sources of cosmic rays,
explain which particles should be observed, and provide pointers to 
publicly-available logs from reliable measurement facilities.
We hope that the shared resources and presented methodology will 
become a standard for any future analysis of the relationship between cosmic rays and DRAM errors in the field. 
Also, numerous past studies on DRAM errors can (and should) be revisited and extended with this analysis.

%% file: 60-Related_work.tex
Many recent studies have presented extensive \textbf{quantitative analyses of DRAM errors in the field}.  
%
The studies analyze error rates and correlate them with chip capacity, temperature, utilization, 
aging and DIMM generation~\cite{Schroeder:SIGMETRICS2009}. 
The papers also distinguish between the transient (soft) and non-transient (hard) errors~\cite{Li:USENIX2010, Hwang:ASPLOS2012}, 
and consider the position 
of the DRAM device in the data-center~\cite{Sridharan:SC2013} and the error position in the chip~\cite{Sridharan:SC2012, Sridharan:SC2013, Sridharan:ASPLOS2015, Meza:DSN2015, Zivanovic:MEMSYS2019, Beigi:HPCA2023}.

It is widely accepted that cosmic rays are a plausible cause of DRAM errors in 
datacenters~\cite{Zivanovic:MEMSYS2019, boixaderas2020cost, boixaderas2024rl, Schroeder:SIGMETRICS2009, Li:USENIX2010, Giurgiu:Middleware2017, Gupta:SC2017, das2021systemic, Levy:SC2018, Du:ICCD2021, li2022correctable}.  
Also, cosmic rays are frequently used as an explanation for some aspects of the observed DRAM error 
behaviors~\cite{Schroeder:TDSC2010, bautista2016unprotected, das2021systemic, Sridharan:SC2013, Sridharan:ASPLOS2015}. 
Schroeder and Gibson~\cite{Schroeder:TDSC2010} analyze error logs of 22~HPC systems that were in production use at Los Alamos National Laboratory between 1996 and 2005.
The authors detect high DRAM error rates, and suggest that 
part of the problem could be high altitude of the cluster, which makes it more exposed to cosmic rays. 
Das et al.~\cite{das2021systemic} analyze the root causes of failures in peta-scale HPC clusters between 2014 and 2016.  
For some failures, the system logs have insufficient information and show no prior symptoms. 
The authors, therefore, suspect that a potential cause of these failures could be solar flares or rare cosmic radiations.   
Sridharan et al.~\cite{Sridharan:SC2013} analyze DRAM and SRAM errors from two supercomputers 
located at Los Alamos National Lab (7320 feet altitude) and Oak Ridge National Lab (850 feet altitude). 
The high-altitude Los Alamos facility reports significantly higher SRAM error rates. 
The authors consider that this finding is sufficient to conclude
that the majority of observed SRAM faults are caused by cosmic ray strikes.  
The authors also suggest that cosmic rays are another possible cause of the elevated fault rates in the higher rack chassis.  
The follow-up study~\cite{Sridharan:ASPLOS2015} from the same HPC facility 
concludes that some (but not all) DRAM devices are susceptible to transient faults from cosmic rays, 
and that this susceptibility highly varies by device manufacturer.  
All these studies consider cosmic rays to be an intuitive hypothetical cause of DRAM errors in the field.  
However, none of them tests this hypothesis. 

A quantitative correlation between DRAM error rates and cosmic rays intensity is performed only by El-Sayed and Schroeder~\cite{el2013reading}. 
The study analyzes error logs from HPC clusters at Los Alamos National Lab 
with the objective to discover the factors that influence the HPC system reliability.  
One out of many factors under study is the cosmic radiation. 
The authors visually analyze monthly probability of a uncorrected DRAM error as a function of the monthly average neutron counts 
reported by a neutron monitor located 400\,km north from the national lab.  
The results suggest that cosmic rays are not associated with higher uncorrected DRAM error rates. Actually, the results suggest the opposite trend --- the higher the neutron counts, the lower the number of errors. The authors find this unexpected and provide one possible explanation for it: while increased rates of cosmic rays might lead to a
higher number of corrupted bits, the types of corruption caused by those events might usually be corrected with the error correction codes. 
This reasoning is aligned with the idea that transient errors typically lead to a single-bit data corruption~\cite{Li:USENIX2010}. 
We agree that contemporary error correcting codes, capable to correct up to 4~corrupted bits in a 72-bit word, 
would probably correct a vast majority of the transient errors caused by an external radiation. 
For this reason, we extend the study of El-Sayed and Schroeder with analysis of the corrected and transient corrected errors. 
Also, apart from the visual correlation, we apply formal statistical tests and analyze the reported correlation coefficients and their statistical significance. 
We also use statistical tests to explore whether the error rates have different distributions during periods of elevated cosmic ray activity, 
e.g. above $90^{th}$ or $99^{th}$ neutron count percentile.
We perform a vast number of tests that cover all combinations of the error categories, time windows and system scopes. 
Finally, we analyze in detail all tests that showed statistical significance. 

\textbf{Machine learning methods predict DRAM errors} based on the error history, 
device characteristics (e.g. manufacturer and technology), 
sensors that monitor the system (e.g. temperature), 
and workload properties (e.g. memory bandwidth utilization)~\cite{Costa:ProactiveMemory-ErrorAvoidance, Baseman:DRAMFaultCharacterization, Baseman:DSN17, Sun:FailurePredictionUsingDL, Du:FailurePredictionUsingOnlineLearning, Nie:HPCA2016, Nie:DSN2018, Giurgiu:Middleware2017, mukhanov2019workload, wang2021workload, boixaderas2020cost, boixaderas2024rl, li2022correctable, zhang2022predicting}. 
To the best of our knowledge, our study is the first one to explore whether considering the neutron count measurements
would enhance these prediction methods.   

\textbf{Physics-based error models} require the simulation of physical processes, nuclear reactions and radiation transport models, that generate a charge capable of impacting the functionality of electronic circuits. This simulation requires complex commercial tools and publicly unavailable circuit design details. Purely physics-based error models are impracticable even for device manufacturers because the simulation (solving coupled differential equations) is typically beyond the reach of available computing power~\cite{Seifert:2010}. Additionally, DRAM error models should consider memory traffic, data values and access patterns~\cite{ziegler1996ibm, Ziegler:1996c, JEDEC:JESD89, JEDEC:JESD89-3}, which can vary greatly depending on the customer and their target applications.

As an alternative, \textbf{semi-empirical models} calibrate the projections of physics-based simulations with error rates from experiments in which devices are irradiated with neutron beams of known flux~\cite{Seifert:2010, Srinivasan:1996, ziegler1996terrestrial, Ziegler:1996c}. Beam experiments, however, also present challenges. The first one is providing consistent conditions among different tests~\cite{ziegler1996ibm, Ziegler:1996c}. Since 2001, these conditions have been standardized by JEDEC~\cite{JEDEC:JESD89, JEDEC:JESD89-3}. Second, the complexity of these experiments limits the number of studies. In the last couple of decades only three studies have reported beam-test results for DRAM and HBM devices~\cite{Borucki:2007, Luza:2021, Sullivan:2021}. Finally, beam experiments may not match production operating conditions~\cite{ziegler1996ibm, Ziegler:1996c}. For example, the spectrum of neutron energies and fluxes can vary significantly between different geographical locations~\cite{Seifert:2010, ziegler1996terrestrial, ziegler1996ibm, Ziegler:1996c}. For this reason, recent JEDEC standards include a flux calculator to determine the relative neutron flux at a particular location and adjust it with respect to a reference location 
(New York City)~\cite{SEU-test}. However, this flux calculation does not account for real-time variations in cosmic-ray intensity, such as those due to local atmospheric pressure or solar and cosmic events~\cite{ziegler1996terrestrial}. In our study, this variation is depicted in Figure~\ref{fig:calma_logs}. Some studies report much higher variations of up to 30\% over just a few days~\cite{Seifert:2010}. Other factors that are difficult to correct include the presence of buildings and whether the system is located on the top floor or in the 
basement~\cite{Seifert:2010}.

Overall, the analysis of DRAM errors induced by cosmic rays presents many challenges. \textbf{In-field studies} are essential because they report device error rates under production conditions: running representative workloads and datasets within a representative hardware and software infrastructure~\cite{Ziegler:1996c, ResilientHPC:2020}. Additionally, due to the lack of detailed physical device models and beam-test results, in-field studies currently serve as the primary source of information on DRAM error rates.

%% file: 70-Conclusions.tex
In this paper we describe the underlying sources of cosmic rays, explain which particles should be observed,
and point to publicly-available logs from reliable measurement facilities. 
Then we analyze cosmic rays and DRAM errors of two 
HPC clusters located in Barcelona, 100\,meters above the sea level. 
We explore correlations between cosmic ray intensity and the number of DRAM errors,
and apply statistical tests to study whether the error rates have different distributions during periods of elevated cosmic
ray activity, e.g. above $90^{th}$ or $99^{th}$ percentile. 
The exploration covers a vast number of tests that include numerous error categories, time windows
and system scopes. We also examine whether machine learning
methods for DRAM error prediction would be enhanced if
they considered the neutron count measurements. 

For the two HPC systems under study, we detect no apparent correlation between DRAM errors and cosmic rays,
and no error rate increment even during periods with the highest cosmic ray intensity. 
We also see that considering neutron counts does not improve DRAM error predictions. 
In addition to this, we apply the presented methodology to verify the findings from previous studies which claim that some DRAM errors are mostly caused by cosmic rays.  
We show that these results are unreliable and the conclusions incorrect. 
Overall, we detect no indications that cosmic rays have any influence on DRAM errors.  
To understand whether the findings are specific to the systems under study, 
the analysis should be repeated on other HPC clusters, especially the ones located on higher altitudes.  
Additionally, the analysis can (and should) be applied to revisit and extend numerous previous studies that use
cosmic rays as a hypothetical explanation for certain aspects of the observed DRAM error behaviors. 
Overall, our study motives, guides and enables a future research on the frequently mentioned, but insufficiently studied 
relationship between cosmic rays and DRAM errors in the field.